\documentclass[aps,graphicx,12pt,showkeys,showpacs]{revtex4}
\usepackage{amsmath}
\usepackage{amscd}
\usepackage{graphicx}
\begin{document}

\title[Short Title]{Atomic quantum state transferring and swapping via quantum Zeno dynamics}

\author{Zhi-Cheng Shi$^{1}$}
\author{Yan Xia$^{1,2, }$\footnote{corresponding author E-mail: xia-208@163.com}}
\author{Jie Song$^{3}$}

\affiliation{$^{1}$Department of Physics, Fuzhou University, Fuzhou
350002, China \\ $^{2}$School of Physics and Optoelectronic
Technology, Dalian University of Technology, Dalian 116024,
China\\$^{3}$Department of Physics, Harbin Institute of Technology,
Harbin 150001, China}

\begin{abstract}In this paper, we first demonstrate how to realize quantum state
transferring (QST) from one atom to another based on quantum Zeno
dynamics. Then, the QST protocol is generalized to realize the
quantum state swapping (QSS) between two arbitrary atoms with the
help of a third one. Furthermore, we also consider the QSS within a
quantum network. The influence of decoherence is analyzed by
numerical calculation. The results demonstrate that the protocols
are robust against cavity decay.

\end{abstract}

\pacs { 03. 67. Hk, 03. 65. Xp }
  \keywords{Quantum state transferring and swapping; Quantum Zeno effect; Cavity quantum electrodynamics}

 \maketitle

\section{INTRODUCTION}
Quantum information processing (QIP) \cite{MAN-CUP2000,HJK-NAT453}
has demonstrated an important development in recent years. Many
protocols for QIP have been proposed in different quantum systems,
such as cavity QED \cite{sbzheng}, trapped-ion systems
\cite{Kielpinski}, quantum-dot systems \cite{bayer}, superconducting
quantum systems \cite{sqs1,sqs2}, and linear optical systems
\cite{los1,los2,los3}. The significant advances in implementing
various protocols for QIP can, in the future, lead to long-distance
quantum communication or creation of a quantum computer.

Quantum information transferring (QIT) from one to another place is
an important goal in the field of quantum information science. So
far, a lot of substantial efforts have been devoted to the field of
QIT
 and much important progresses have been made
\cite{AK-PRL85,AB-PRA70,JFZ-PRA73,HW-PRA76,AB-PRA75,CDF-PRA81,BC-PRA83,PBL-PRA79,JIC-PRL78,ADB-PRL98}.
An important element of many QIP operations is the transfer of a
quantum state from one to another qubit \cite{SB-PRL91}, such as
$|0\rangle_{A}\otimes(a|0\rangle+b|1\rangle)_{B} \rightarrow
(a|0\rangle+b|1\rangle)_{A}|0\rangle_{B}$, where
$|a|^{2}+|b|^{2}=1$, $A$ and $B$ denote atom $A$ and atom $B$,
respectively. From this expression, we can notice that the problem
about the quantum information transfer (QIT) can be reduced to the
issue of quantum state transfer (QST) in some sense if the quantum
information is encoded in the states of atoms. There are many
methods that can implement the QST, such as making use of quantum
teleportation original proposed by Bennett \emph{et al.}
\cite{CHB-PRL70}, which has been experimentally realized in optical
and liquid-state nuclear magnetic resonance (NMR) systems
\cite{DB-NAT390,DB-PRL80,MAN-NAT396}. In addition, atomic systems in
the context of cavity QED are suitable to act as qubits because
moderate internal electronic states can coherently store information
over very long time scale. In 2010, Yang \cite{yang} have proposed a
way of implementing QST with two superconducting flux qubits by
coupling them to a resonator. This proposal does not require
adjustment of the level spacings or uniformity in the device
parameters.

On the other hand, Facchi \emph{et al.}
\cite{11PLA275,12PRA65,13PIO42} found the quantum Zeno dynamics
which is a broader formulation of the quantum Zeno effect
\cite{09JMP18}, since the system will evolve away from its initial
state and remains in the Zeno subspace determined by the measurement
when frequently projected onto a multi-dimensional subspace and the
quantum Zeno effect can be reformulated in terms of a continuous
coupling to obtain the same effect without making use of von
Neumann's projections and non-unitary dynamics. Until now, the new
finding has enlightened numerous schemes to implement quantum
computation and prepare quantum entanglement
\cite{PJ-PRL89,KP-PRA69,XQ-PLA374,SZ-PBA44,14PRA77,XQS-EPL90,XQS-NJP12,ALW-OC284,15PRA83}.

 In this paper, we will first
demonstrate how to implement the QST based on quantum Zeno dynamics.
Then, the QST protocol will be generalized to realize the quantum
state swapping (QSS) between two arbitrary atom. Moreover, the QSS
within in a quantum network will be also considered in present
paper. The setup is composed of cavity QEDs, optical fibers, and
$\Lambda$-type atoms, which make the scheme feasible with the
current technology. We will show that the protocols are robust
against cavity decay and a relatively high fidelity can be obtained
even in the presence of atomic spontaneous emission and optical
fiber decay.

Before elaborating on our protocols, we first give an elementary
introduction to the quantum Zeno dynamics induced by continuous
coupling \cite{PF-JPC196}. We assume that a system whose dynamical
evolution governed by a generical Hamiltonian $H_{K}=H+KH_{c},$
where $H$ is the Hamiltonian of the system to be investigated and
$H_{c}$ can be regarded as an additional interaction Hamiltonian,
which perform the ``measurement". \emph{K} is a coupling constant.
For the infinitely strong coupling limit $K\rightarrow \infty$, the
system is dominated by the limiting evolution operator
$U_{K}(t)=\lim_{K\rightarrow \infty}
\exp(-\emph{i}KH_{c}t)\mathcal{U}(t)$, where $\mathcal{U}(t)=
\exp(-\emph{i}H_{z}t)$ \cite{PF-PRL89}. $H_{z}=\sum_{n}P_{n}HP_{n}$
is viewed as the Zeno Hamiltonian and $P_{n}$ is the eigenprojection
of $H_{c}$ correspondence to the eigenvalue $\eta_{n}$
($H_{c}=\sum_{n}\eta_{n}P_{n}, (\eta_{n} \neq \eta_{m}, \texttt{for}
~n \neq m)$). Therefore, the limiting evolution operator of the
system can be depicted as $U_{K}(t) \sim
\exp(-\emph{i}KH_{c}t)\mathcal {U}(t) =
\exp(-\emph{i}\sum_{n}K\eta_{n}P_{n}t+P_{n}HP_{n}t)$. Thus, we can
derive the expression of the effective Hamiltonian of the system:
$H_{eff}=\sum_{n}(K\eta_{n}P_{n}+P_{n}HP_{n}),$ which is an
important result to the following works that are based on.

\section{The QST From atom 2 to atom 1}

As depicted in Fig. 1, we consider that two identical atoms (1, 2),
which have one excited state $|e\rangle$ and two ground states
$|0\rangle$ and $|1\rangle$ with a $\Lambda$-type three-level
configuration, are trapped in distant cavities ($c_1$, $c_2$)
connected by one optical fiber {$f$}, respectively. Suppose that the
transition $|e\rangle_k \leftrightarrow |0\rangle_k$ (\emph{k}=1,2)
is resonantly coupled to the cavity mode with the coupling strength
$g_{k}$ while the transition $|e\rangle_k\leftrightarrow
|1\rangle_k$ is resonantly driven by a classical laser field with
the Rabi frequency $\Omega_{k}$. Let $L$ be the length of fiber, $C$
be the speed of light, and $\bar{\nu}$ be the decay rate of the
cavity fields into a continuum of fiber modes. The length $L$ of the
fiber means a quantization of the modes of the fiber with frequency
spacing given by $2\pi C/L$. Then we can have that the number of
modes which would significantly interact with the cavities¡¯ modes
is of the order of $n=L\bar{\nu}/2\pi C$ \cite{TP-PRL79}. In the
short fiber limit $L\bar{\nu}/2\pi C$ \cite{AS-PRL96}, which focuses
on the case $n\leq 1$, only one fiber mode is essentially excited
and coupled to the cavity mode with coupling strength $\lambda$.
Notice that such a regime applies in most realistic experimental
situations: for instance, $L\leq 1$ m and $\bar{\nu}\simeq 1$ GHz
(natural units are adopted with $h=1$) are in the proper range. In
the interaction picture, the Hamiltonian for the whole system can be
written as ($\hbar=1$)
\begin{eqnarray}\label{01}
H_{tot}&=& H_{l}+H_{c}+H_{cf},\nonumber\\
H_{l}&=&\sum_{k=1}^{2}\Omega_{k}(|e\rangle_{k}\langle 1|+|1\rangle_{k} \langle e|),\nonumber\\
H_{c}&=&\sum_{k=1}^{2}g_{k}(a_{k}|e\rangle_k\langle 0|+a_{k}^{\dag}|0\rangle_k\langle e|),\nonumber\\
H_{cf}&=&\lambda b(a_{1}^{\dag}+a_{2}^{\dag})+H.c.,
\end{eqnarray}
where $a_{k}^\dagger$ and $a_{k}$ are the creation and annihilation
operators for the \emph{k}th cavity mode and $b^\dagger$ and $b$ are
the creation and annihilation operators for the fiber mode. We
assumed $g_{k}$ = $g\in R$ for simplicity.

If the initial state of the system is
$|0\rangle_{1}|1\rangle_{2}|0\rangle_{c_{2}}|0\rangle_{f}|0\rangle_{c_{1}}$,
it will evolve in a closed subspace spanned by $\{|\phi_{1}\rangle,
|\phi_{2}\rangle, |\phi_{3}\rangle, |\phi_{4}\rangle,
|\phi_{5}\rangle, |\phi_{6}\rangle, |\phi_{7}\rangle\}$, where
\begin{eqnarray}\label{02}
|\phi_{1}\rangle&=&|0\rangle_{1}|1\rangle_{2}|0\rangle_{c_{2}}|0\rangle_{f}|0\rangle_{c_{1}},\nonumber\\
|\phi_{2}\rangle&=&|0\rangle_{1}|e\rangle_{2}|0\rangle_{c_{2}}|0\rangle_{f}|0\rangle_{c_{1}},\nonumber\\
|\phi_{3}\rangle&=&|0\rangle_{1}|0\rangle_{2}|1\rangle_{c_{2}}|0\rangle_{f}|0\rangle_{c_{1}},\nonumber\\
|\phi_{4}\rangle&=&|0\rangle_{1}|0\rangle_{2}|0\rangle_{c_{2}}|1\rangle_{f}|0\rangle_{c_{1}},\nonumber\\
|\phi_{5}\rangle&=&|0\rangle_{1}|0\rangle_{2}|0\rangle_{c_{2}}|0\rangle_{f}|1\rangle_{c_{1}},\nonumber\\
|\phi_{6}\rangle&=&|e\rangle_{1}|0\rangle_{2}|0\rangle_{c_{2}}|0\rangle_{f}|0\rangle_{c_{1}},\nonumber\\
|\phi_{7}\rangle&=&|1\rangle_{1}|0\rangle_{2}|0\rangle_{c_{2}}|0\rangle_{f}|0\rangle_{c_{1}}.
\end{eqnarray}
The subscripts $1$, $2$, $c_1$, $f$ and $c_2$ represent the atom 1,
atom 2, cavity 1, optical fiber and cavity 2, respectively.

On the condition that $\Omega_{1}, \Omega_{2}\ll g, \lambda$, the
Hilbert subspace is split into five invariant Zeno subspaces
\cite{PF-JPC196,PF-PRL89}:
\begin{eqnarray}\label{03}
H_{P_{1}}= \{ |\phi_{1}\rangle, |\phi_{7}\rangle,
|\varphi_{1}\rangle \}, H_{P_{2}}= \{ |\varphi_{2}\rangle \},
H_{P_{3}}= \{ |\varphi_{3}\rangle \}, H_{P_{4}}= \{
|\varphi_{4}\rangle \}, H_{P_{5}}= \{ |\varphi_{5}\rangle \},
\end{eqnarray}
where
\begin{eqnarray}\label{04}
|\varphi_{1}\rangle&=&\frac{1}{\sqrt{2\lambda^{2}+g^{2}}}(\lambda|\phi_{2}\rangle-g|\phi_{4}\rangle+\lambda|\phi_{6}\rangle),\nonumber\\
|\varphi_{2}\rangle&=&\frac{1}{2}(-|\phi_{2}\rangle+|\phi_{3}\rangle-|\phi_{5}\rangle+|\phi_{6}\rangle),\nonumber\\
|\varphi_{3}\rangle&=&\frac{1}{2}(-|\phi_{2}\rangle-|\phi_{3}\rangle+|\phi_{5}\rangle+|\phi_{6}\rangle),\nonumber\\
|\varphi_{4}\rangle&=&\frac{1}{2\sqrt{2\lambda^{2}+g^{2}}}(g|\phi_{2}\rangle-\sqrt{2\lambda^{2}+g^{2}}|\phi_{3}\rangle
+2\lambda|\phi_{4}\rangle-\sqrt{2\lambda^{2}+g^{2}}|\phi_{5}\rangle+g|\phi_{6}\rangle),\nonumber\\
|\varphi_{5}\rangle&=&\frac{1}{2\sqrt{2\lambda^{2}+g^{2}}}(g|\phi_{2}\rangle+\sqrt{2\lambda^{2}+g^{2}}|\phi_{3}\rangle
+2\lambda|\phi_{4}\rangle+\sqrt{2\lambda^{2}+g^{2}}|\phi_{5}\rangle+g|\phi_{6}\rangle),
\end{eqnarray}
corresponding to eigenvalues $\eta_{1}=0, \eta_{2}=-g, \eta_{3}=g,
\eta_{2}=-\sqrt{2\lambda^{2}+g^{2}},
\eta_{3}=\sqrt{2\lambda^{2}+g^{2}}$ with the projections
\begin{eqnarray}\label{05}
P_{n}=\sum_{j}|\beta_{n,j}\rangle \langle \beta_{n,j}|,
(|\beta_{n,j}\rangle\in H_{P_{n}}).
\end{eqnarray}
Therefore, the Hamiltonian of the current system is approximately
governed by
\begin{eqnarray}\label{06}
H_{total}&\cong& \sum_{n}(\eta_{n}P_{n}+P_{n}H_{laser}P_{n})
\cr\cr&=&-g|\varphi_{2}\rangle\langle
\varphi_{2}|+g|\varphi_{3}\rangle\langle
\varphi_{3}|-\sqrt{2\lambda^{2}+g^{2}}|\varphi_{4}\rangle\langle
\varphi_{4}|+\sqrt{2\lambda^{2}+g^{2}}|\varphi_{5}\rangle\langle
\varphi_{5}|{}
\nonumber\\
& & {}
+\frac{\lambda}{\sqrt{2\lambda^{2}+g^{2}}}(\Omega_{1}|\phi_{1}\rangle\langle
\varphi_{1}|+\Omega_{2}|\phi_{7}\rangle\langle \varphi_{1}|)+H.c..
\end{eqnarray}
As the initial state is
$|0\rangle_{1}|1\rangle_{2}|0\rangle_{c_{2}}|0\rangle_{f}|0\rangle_{c_{1}}$,
thus the effective Hamiltonian of system reduces to
\begin{eqnarray}\label{07}
H_{eff}&=&\frac{\lambda}{\sqrt{2\lambda^{2}+g^{2}}}(\Omega_{1}|\phi_{1}\rangle\langle
\varphi_{1}|+\Omega_{2}|\phi_{7}\rangle\langle \varphi_{1}|)+H.c..
\end{eqnarray}

On the other hand, it is easily checked that the evolution of
initial state
$|0\rangle_{1}|0\rangle_{2}|0\rangle_{c_{2}}|0\rangle_{f}|0\rangle_{c_{1}}$
is frozen due to
$H_{tot}|0\rangle_{1}|0\rangle_{2}|0\rangle_{c_{2}}|0\rangle_{f}|0\rangle_{c_{1}}=0$.

As the initial state of the whole system is
$|\Phi(0)\rangle=|0\rangle_{1}\otimes(a|0\rangle+b|1\rangle)_{2}
\otimes|0\rangle_{c_{2}}|0\rangle_{f}|0\rangle_{c_{1}}$, where
$|a|^{2}+|b|^{2}=1$, it will evolve with respect to the effective
Hamiltonian in Eq. (7). Set $\Omega_{1}=-\Omega_{2}=\Omega\in R.$
For an interaction time $t$, the final state of the system becomes
\begin{eqnarray}\label{08}
|\Phi(t)\rangle
&=&a|0\rangle_{1}|0\rangle_{2}|0\rangle_{c_{2}}|0\rangle_{f}|0\rangle_{c_{1}}+b[
\frac{1}{2}(1+\cos{\frac{\sqrt{2}\lambda\Omega
t}{\sqrt{2\lambda^{2}+g^{2}}}})|\phi_{1}\rangle+\frac{1}{2}(1-\cos{\frac{\sqrt{2}\lambda\Omega
t}{\sqrt{2\lambda^{2}+g^{2}}}})|\phi_{7}\rangle{}
\nonumber\\
& & {}-i\frac{\sqrt{2}}{2}\sin{\frac{\sqrt{2}\lambda\Omega
t}{\sqrt{2\lambda^{2}+g^{2}}}}|\varphi_{1}\rangle].
\end{eqnarray}
By selecting the interaction time to satisfy
$\frac{\sqrt{2}\lambda\Omega t}{\sqrt{2\lambda^{2}+g^{2}}}=\pi$, one
will obtain
\begin{eqnarray}\label{09}
|\Phi(\frac{\sqrt{2\lambda^{2}+g^{2}}\pi}{\sqrt{2}\lambda\Omega})\rangle&=&a|0\rangle_{1}|0\rangle_{2}|0\rangle_{c_{2}}|0\rangle_{f}|0\rangle_{c_{1}}
+b|\phi_{7}\rangle\cr\cr&=&a|0\rangle_{1}|0\rangle_{2}|0\rangle_{c_{2}}|0\rangle_{f}|0\rangle_{c_{1}}+b|1\rangle_{1}|0\rangle_{2}|0\rangle_{c_{2}}|0\rangle_{f}|0\rangle_{c_{1}}
\cr\cr&=&(a|0\rangle+b|1\rangle)_{1}\otimes
|0\rangle_{2}\otimes|0\rangle_{c_{2}}|0\rangle_{f}|0\rangle_{c_{1}},
\end{eqnarray}
where the QST from atom 2 to atom 1 has been realized.

\section{The QSS between atoms 2 and 3 with the help of atom 1}

Now, we will demonstrate that how to swap the quantum states between
atom 2 and atom 3 with the help of the auxiliary atom 1, as shown in
Fig. 2. Assume that the initial arbitrary states of atom 2 and atom
3 are $(a|0\rangle+b|1\rangle)_{2}$ and
$(c|0\rangle+d|1\rangle)_{3}$ ($|a|^{2}+|b|^{2}=1,
|c|^{2}+|d|^{2}=1$), respectively. In addition, all the optical
switches are closed in the initial time. During the swapping
operations, we will introduce a auxiliary atom 1 with the initial
state $|0\rangle$.

\subsection{The QST From Atom 2 To Atom 1}

First, we turn on the optical switches 1 and 2 to let the optical
fiber mediate the cavities 1 and 2. The initial state of the system
is $|\Phi(0)\rangle=|0\rangle_{1}\otimes(a|0\rangle+b|1\rangle)_{2}
\otimes|0\rangle_{c_{2}}|0\rangle_{f}|0\rangle_{c_{1}}$. A analogue
analysis is utilized with the Eq. (2) - Eq. (8), then set Rabi
frequency $\Omega_{1}=-\Omega_{2}=\Omega\in R$ and an interaction
time $\frac{\sqrt{2\lambda^{2}+g^{2}}\pi}{\sqrt{2}\lambda\Omega}$.
We will realize the QST from atom 2 to atom 1. Next, we turn off the
optical switches 1 and 2 to inhibit the interaction between atom 1
and atom 2. As a consequence, the quantum state of atom 1 becomes
$(a|0\rangle+b|1\rangle)_{1}$ while the final state of atom 2
becomes $|0\rangle_{2}$.

\subsection{The QST From Atom 3 To Atom 2}

Then, we turn on the optical switches 2 and 3 to let the optical
fiber mediate the cavities 2 and 3. The initial state of the system
is
$|\Phi^{'}(0)\rangle=|0\rangle_{2}\otimes(c|0\rangle+d|1\rangle)_{3}
\otimes|0\rangle_{c_{2}}|0\rangle_{f}|0\rangle_{c_{1}}$. A analogue
analysis is utilized with the Eq. (2) - Eq. (8), then set Rabi
frequency $\Omega_{2}=-\Omega_{3}=\Omega\in R$ and an interaction
time $\frac{\sqrt{2\lambda^{2}+g^{2}}\pi}{\sqrt{2}\lambda\Omega}$.
We will realize the QST from atom 3 to atom 2. Next, we turn off the
optical switches 2 and 3 to inhibit the interaction between atom 2
and atom 3. As a consequence, the quantum state of atom 2 becomes
$(c|0\rangle+d|1\rangle)_{2}$ while the final state of atom 3
becomes $|0\rangle_{3}$.

\subsection{The QST From Atom 1 To Atom 3}

Finally, we turn on the optical switches 1 and 3 to let the optical
fiber mediate the cavities 1 and 3. The initial state of the system
is$|\Phi^{''}(0)\rangle=|0\rangle_{3}\otimes(a|0\rangle+b|1\rangle)_{1}
\otimes|0\rangle_{c_{2}}|0\rangle_{f}|0\rangle_{c_{1}}$. A analogue
analysis is utilized with the Eq. (2) - Eq. (8), then set Rabi
frequency $\Omega_{3}=-\Omega_{1}=\Omega\in R$ and an interaction
time $\frac{\sqrt{2\lambda^{2}+g^{2}}\pi}{\sqrt{2}\lambda\Omega}$.
We will realize the QST from atom 1 to atom 3. Next, we turn off the
optical switches 1 and 3 to inhibit the interaction between atom 1
and atom 3. As a consequence, the quantum state of atom 3 becomes
$(a|0\rangle+b|1\rangle)_{3}$ while the final state of the auxiliary
atom 1 becomes $|0\rangle_{1}$.

After above operations, we have realized the QSS between atom 2 and
atom 3, which become $(c|0\rangle+d|1\rangle)_{2}$ and
$(a|0\rangle+b|1\rangle)_{3}$ now, while the final state of
auxiliary atom 1 remains $|0\rangle_{1}$.

\section{The QSS For Two Arbitrary Atoms Among $N$ Atoms Within The Quantum Network}

From above analysis, we can find that the state of auxiliary atom 1
remains unchange after two other atoms 2 and 3 have realized the
QSS. Thus, it provides a scalable way to realize the QSS for two
arbitrary atoms among $N$ atoms within the quantum network.

As shown in Fig. 3, $N$ atoms are trapped in $N$ separate cavities,
respectively. The $N$ cavities are connected by the fibers and $N$
optical switches within the quantum network. Now we briefly
demonstrate how to realize one QSS. For example, if we want to swap
the arbitrary two atomic quantum state within the quantum network,
for example, atom 4 and arbitrary atom $N$, the first step we must
do is to turn off all the optical switches in quantum network. Next,
we turn on the optical switches 1 and 4 to realize the QST from atom
4 to atom 1. Then we turn off the optical switch 1 and turn on the
optical switch $N$ to realize the QST from atom $N$ to atom 4.
Finally, we turn off the optical switch 4 and turn on the optical
switch 1 to realize the QST from atom 1 to atom $N$. Until now, we
have realize the QSS between atom 4 and atom $N$ and turn off the
optical switches 1 and $N$.


\section{NUMERICAL ANALYSIS AND CONCLUSIONS}

All the above results are based on the condition that $\Omega_{1},
\Omega_{2}\ll g, \lambda$. Thus we shall analyze the influence of
the ratio $\Omega/g$ on the fidelity of QST. On the other hand, the
ratio $\lambda/g$ will also affect the fidelity of QST
\cite{AS-PRL96}. We depict the relation between the fidelity of QST
and the ratio $\lambda/g$ and $\Omega/g$ by numerical calculation in
the FIG. 4. Obviously, the smaller $\Omega$ we set, the better
behavior we will get. However, small $\Omega$ implies that long
operation times should be required. We can also see that the
fidelity is above 96\% even though the ratio $\lambda/g=0.1$. Thus
the large cavity-fiber coupling is not necessary needed in an
experiment.

As we can see from the above analysis, the time evolution of the
initial state
$|0\rangle_{1}|0\rangle_{2}|0\rangle_{c_{2}}|0\rangle_{f}|0\rangle_{c_{1}}$
will freeze during the operations. Thus it will transfer with 100\%.
the only factor that will affect the fidelity of QST is the time
evolution of the initial state
$|0\rangle_{1}|1\rangle_{2}|0\rangle_{c_{2}}|0\rangle_{f}|0\rangle_{c_{1}}$.
Therefore we will emphasize on discussion about the fidelity of QST
in the presence of the decoherence induced by cavity decay, optical
fiber decay, and atomic spontaneous emission while the initial state
is
$|0\rangle_{1}|1\rangle_{2}|0\rangle_{c_{2}}|0\rangle_{f}|0\rangle_{c_{1}}$
as follows. When we consider about the decoherence, the master
equation of motion for the density matrix of the whole system can be
express as
\begin{eqnarray}\label{10}
\dot{\rho}&=&-i[H_{tot},\rho]-\sum_{j=1}^{2}\frac{\kappa_{j}}{2}(a_{j}^{\dag}a_{j}\rho-2a_{j}\rho
a_{j}^{\dag}+\rho
a_{j}^{\dag}a_{j})-\frac{\kappa_{f}}{2}(b^{\dag}b\rho-2b\rho
b^{\dag}+\rho b^{\dag}b){}
\nonumber\\
& &
{}-\sum_{k=1}^{2}\sum_{m=0}^{1}\frac{\Gamma_{em}^{k}}{2}(\sigma_{em}^{k}\sigma_{me}^{k}\rho
-2\sigma_{me}^{k}\rho\sigma_{em}^{k}+\rho\sigma_{em}^{k}\sigma_{me}^{k}),
\end{eqnarray}
where $\Gamma_{em}^{k}$ is the spontaneous emission rate of the
$k$th atom from the excited state $|e\rangle$ to the ground state
$|m\rangle$ ($m=0,1$). $\kappa_{j}$ is the decay rate of the $j$th
cavity mode and $\kappa_{f}$ is the decay rate of the optical fiber
mode between two cavities, such as the cavity 1 and cavity 2. For
the sake of simplicity, we assume $\Gamma_{em}^{k}=\Gamma/2$ and
$\kappa_{j}=\kappa$. In Fig. 5 (Fig. 6), we plot the relation of the
fidelity $F$ versus $\kappa/g$ and $\Gamma/g$ ($\kappa_{f}/\lambda$)
by solving the master equation numerically. One can find from Fig. 5
(Fig. 6) that with the increasing of cavity decay and atomic
spontaneous emission (optical fiber decay), the fidelity $F$ of the
QST will decrease. In addition, the results indicate that the QST is
robust against the decay of cavity, since for a large cavity decay
$\kappa/g=0.1$, $\Gamma/g=0$ and $\kappa_{f}/\lambda=0$, the
fidelity is still about 97.21\%. Therefore it can be considered as a
decoherence-free QST with respect to cavity decay. The dominant
decoherence is the atomic spontaneous emission and the optical fiber
decay due to the excited states and the state of fiber with one
photon are included during the evolution. However, the effect of
optical fiber decay is weaker than the effect of atomic spontaneous
emission, which we can account for in this way that the population
probability for one photon in fiber is nearly one-half of the
population probability for excited atoms while $g$ and $\lambda$ are
kept in the same magnitude.

Finally we bring forward the basic elements that may be candidates
for the intended experiment. The requirements of our protocols are
$\Lambda$-type three-level configuration atoms and resonant cavities
connected by optical fibers. The atomic configuration involved in
our proposal can be achieved with a cesium. The state $|0\rangle$
corresponds to $F=3, m=2$ hyperfine state of $6^{2}S_{1/2}$
electronic ground state, the state $|1\rangle$ corresponds to $F=4,
m=4$ hyperfine state of $6^{2}S_{1/2}$ electronic ground state, and
the excited state $|e\rangle$ corresponds to $F=3, m=3$ hyperfine
state of $6^{2}P_{1/2}$ electronic ground state, respectively. In
recent experiments \cite{SM-PRA71,JR-PRA67}, it is achievable with
the parameters $\lambda =2\pi\times750$ MHz, $\Gamma =
2\pi\times2.62$ MHz, $\kappa = 2\pi\times3.5$ MHz in an optical
cavity with the wavelength in the region 630 - 850 nm. A
near-perfect fiber-cavity coupling with an efficiency larger than
99.9\% can be realized using fiber-taper coupling to high-Q silica
microspheres \cite{KJG-IEEE40}. The optical fiber decay at a 852 nm
wavelength is about 2.2 dB/km \cite{FD-PRA75}, which corresponds to
the fiber decay rate 0.152 MHz, lower than the cavity decay rate. By
substituting these typical parameters into Eq. (10), we will obtain
a high fidelity about 97.54\%, which shows the QST in our protocols
are relative robust against realistic one.

In summary, we have proposed a set of protocols for quantum state
transferring and swapping based on quantum Zeno dynamics. The
protocols are robust against cavity decay since it keeps in a closed
subspace without exciting the cavity field during the whole system
evolution. In addition, we have also discussed the influence of
atomic spontaneous emission and optical fiber decay by a
straightforward numerical calculation. The results demonstrate that
a relatively high fidelity can be obtained even in the presence of
atomic spontaneous emission and fiber decay. Therefore, we hope that
it may be possible to realize it in this paper with the current
experimental technology.

\acknowledgements

This work is supported by the funds from Education Department of
Fujian Province of China under Grant No. JB08010, No. JA10009 and
No. JA10039, the National Natural Science Foundation of Fujian
Province of China under Grant No. 2009J06002 and No. 2010J01006, the
National Natural Science Foundation of China under Grant No.
11047122, No. 11105030 and No. 10974028, Doctoral Foundation of the
Ministry of Education of China under Grant No. 20093514110009, and
China Postdoctoral Science Foundation under Grant No. 20100471450.

\newpage

FIG. 1. The experimental setup for realizing the QST from atom 2 to
 atom 1. Those atoms have the identical $\Lambda$-type three-level
configuration.

FIG. 2. The experimental setup for realizing the QSS between atom 2
and atom 3 while atom 1 is an auxiliary atom. The cavities are
connected by optical fibers. The optical switches 1, 2 and 3 can
control two cavities whether have interaction or not.

FIG. 3. The experimental setup for realizing the QSS for two
arbitrary atom among $N$ atoms in the quantum network.

FIG. 4. The fidelity $F$ of QST as a function of the ratio
$\lambda/g$ and $\Omega/g$.

FIG. 5. The fidelity $F$ of QST as a function of cavity decay
$\kappa/g$ and atomic spontaneous emission $\Gamma/g$ in the case of
$\Omega_{1}=0.1g$ and $\kappa_{f}/\lambda=0$.

FIG. 6. The fidelity $F$ of QST as a function of cavity decay
$\kappa/g$ and optical fiber decay $\kappa_{f}/\lambda$ in the case
of $\Omega_{1}=0.1g$, $\Gamma/g=0$ and $g=\lambda$.

\newpage

\begin {figure}
\scalebox{0.8}{\includegraphics {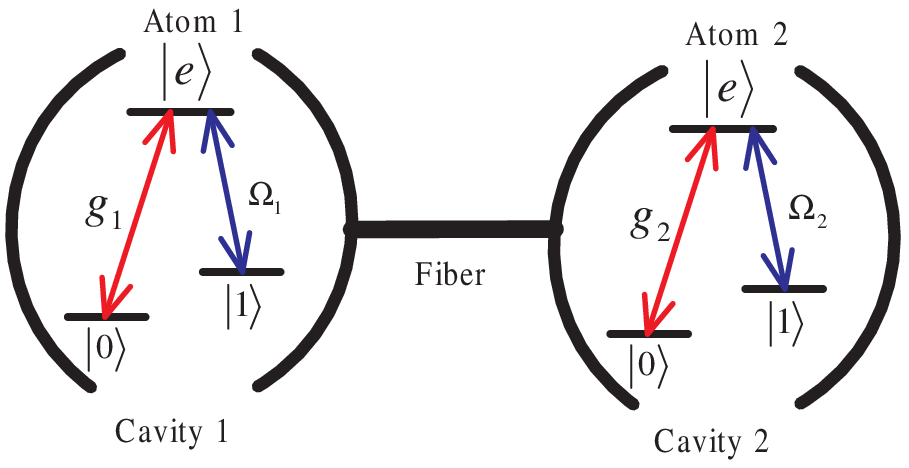}} \caption{} \label
{FIG.1}
\end{figure}

\begin {figure}
\scalebox{0.8}{\includegraphics {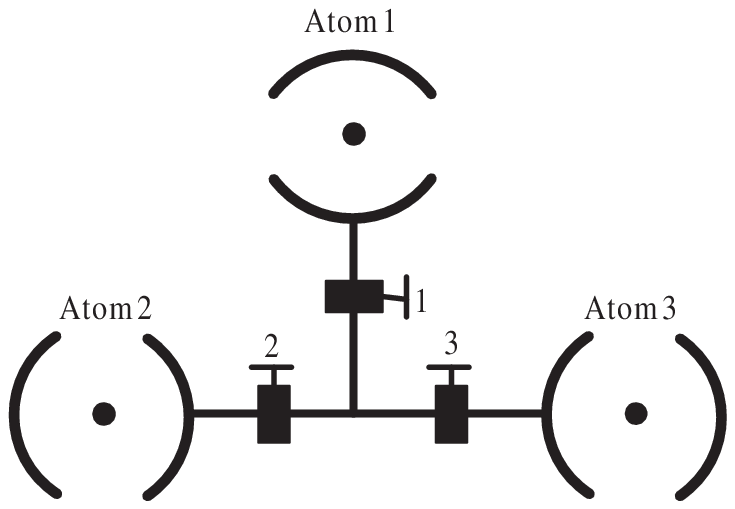}} \caption{} \label
{FIG.2}
\end{figure}

\begin {figure}
\scalebox{0.8}{\includegraphics {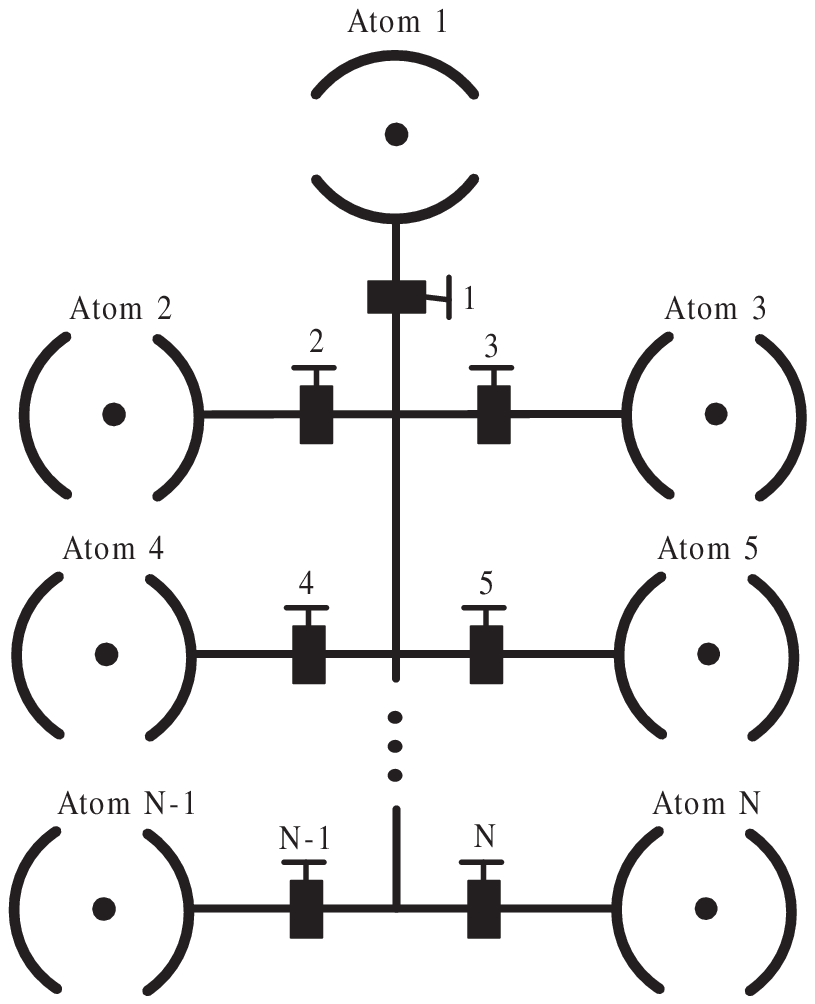}} \caption{} \label
{FIG.3}
\end{figure}

\begin {figure}
\scalebox{0.8}{\includegraphics {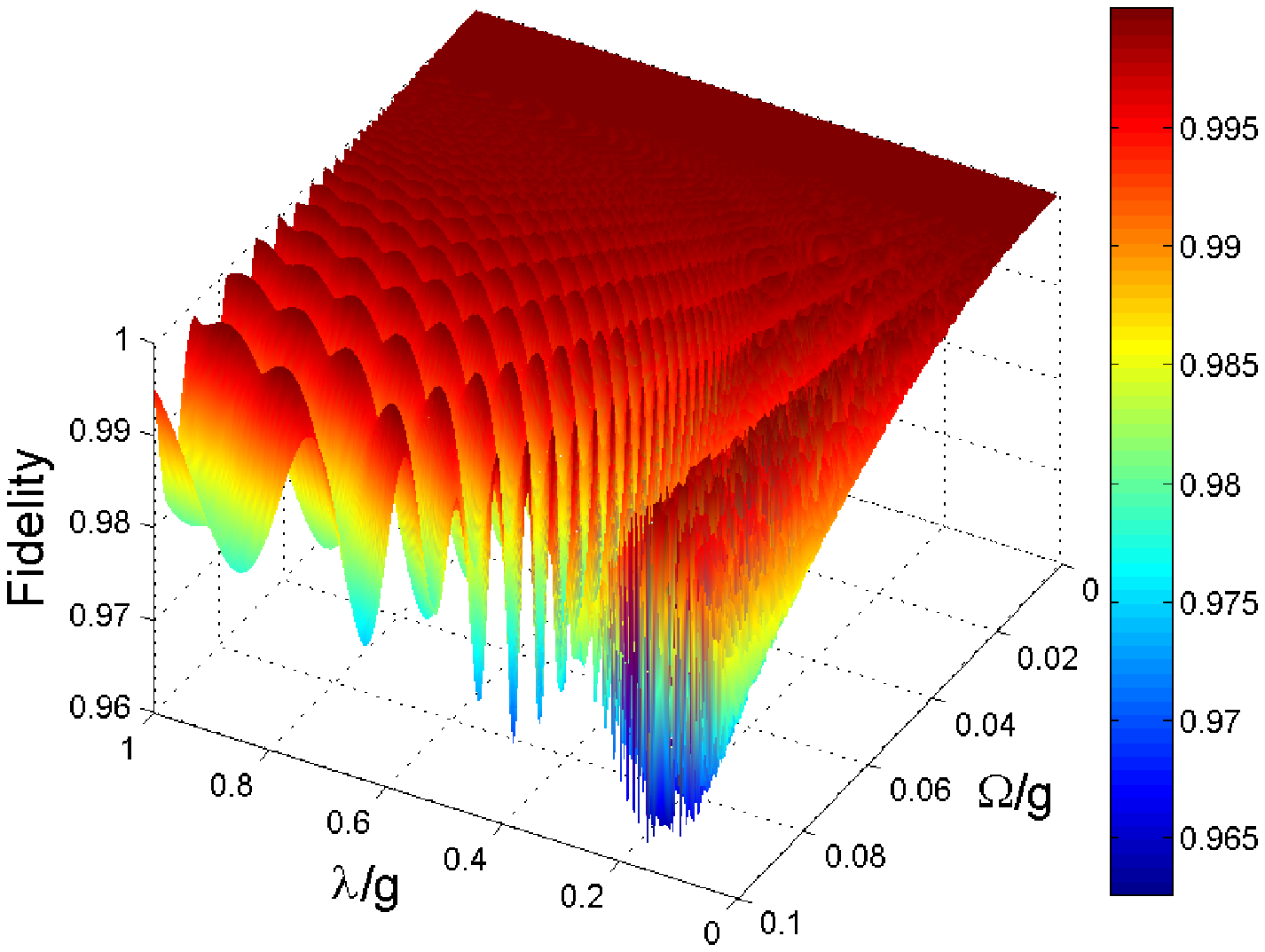}} \caption{} \label
{FIG.4}
\end{figure}

\begin {figure}
\scalebox{0.8}{\includegraphics {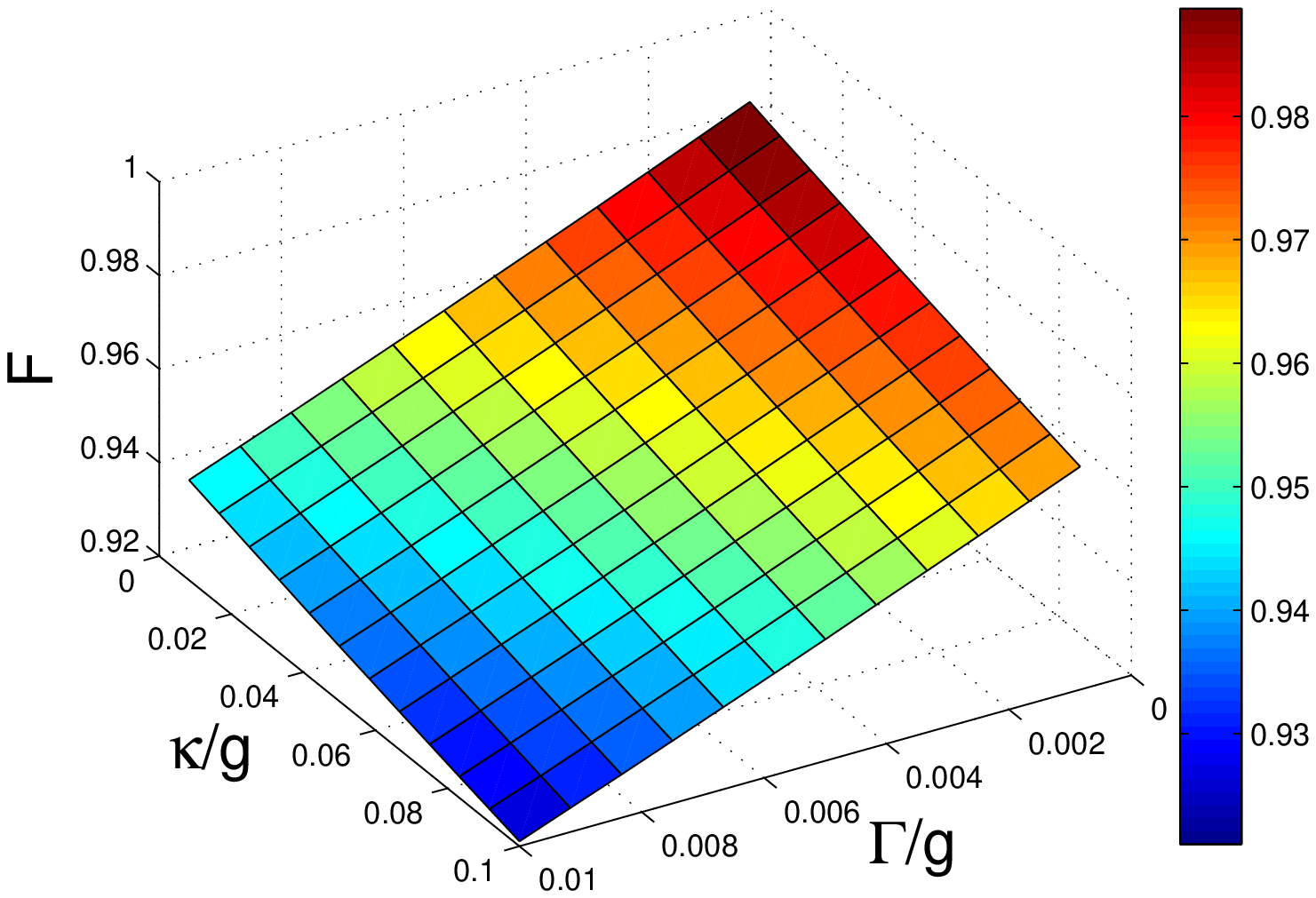}} \caption{} \label
{FIG.5}
\end{figure}

\begin {figure}
\scalebox{0.8}{\includegraphics {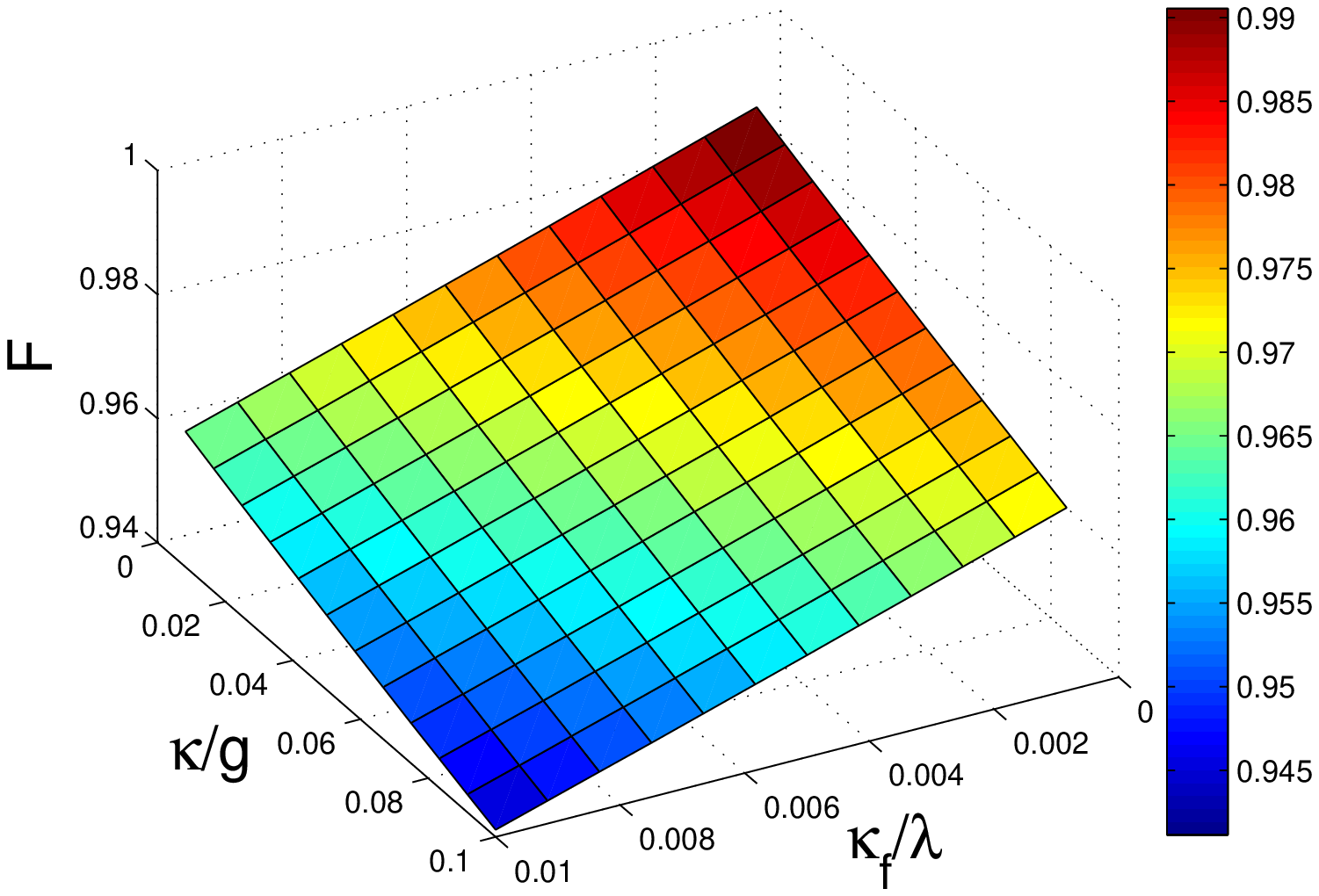}} \caption{} \label
{FIG.6}
\end{figure}


\begin{thebibliography}{999}


\bibitem{MAN-CUP2000}M.~A.~Nielsen and I.~L.~Chuang, Quantum Computation
and Quantum Information (Cambridge University Press, Cambridge,
2000).

\bibitem{HJK-NAT453}H.~J.~Kimble, ``The quantum internet,''
 Nature (London) \textbf{453}, 1023-1030 (2008).

\bibitem{sbzheng}S.~B.~Zheng and G.~C.~Guo, ``Efficient scheme for two-atom entanglement and quantum information
 processing in cavity QED,'' Phys. Rev. Lett. \textbf{85}, 2392-2395 (2000).

\bibitem{Kielpinski}D.~Kielpinski, C.~Monroe, and D.~J.~Wineland, ``Architecture for a large-scale
ion-trap quantum computer,'' Nature (London) \textbf{417}, 709-711
(2002).

\bibitem{bayer}M.~Bayer, P.~Hawrylak, K.~Hinzer, S.~Fafard, M.~Korkusinski,
Z.~R.~Wasilewski, O.~Stern, and A.~Forchel, ``Coupling and
entangling of quantum states in quantum dot molecules,'' Science
\textbf{291}, 451-453 (2001).

\bibitem{sqs1}J.~Q.~You and F.~Nori, ``Quantum information processing with superconducting
qubits in a microwave field,'' Phys. Rev. B \textbf{68}, 064509 (2003).

\bibitem{sqs2}Z.~R.~Lin, G.~P.~Guo, T.~Tu, F.~Y.~Zhu, and G.~C.~Guo, ``Generation of quantum-dot cluster
 states with a superconducting transmission line resonator,'' Phys. Rev.
Lett. \textbf{101}, 230501 (2008).

\bibitem{los1}E.~Knill, R.~Laflamme, and G.~J.~Milburn, ``A scheme for efficient quantum computation with linear
optics,'' Nature (London) \textbf{409}, 46-52 (2001).

\bibitem{los2}J.~W.~Pan, S.~Gasparoni, R.~Ursin, G.~Weihs,
and A.~Zeilinger, ``Experimental entanglement purification of
arbitrary unknown states,'' Nature (London) \textbf{423}, 417-422
(2003).

\bibitem{los3}Y.~Xia, J.~Song, and H.~S.~Song, ``Linear optical protocol for preparation of $N$-photon
 Greenberger-Horne-Zeilinger state with conventional photon detectors,'' Appl. Phys. Lett. \textbf{92}, 021127
(2008); J.~Song, Y.~Xia, an H.~S.~Song, ``Quantum nodes for W-state
generation in noisy channels,'' Phys. Rev. A \textbf{78}, 024302
(2008).









\bibitem{AK-PRL85}A.~Kuzmich and E.~S.~Polzik, ``Atomic quantum state teleportation
 and swapping,'' Phys. Rev. Lett. \textbf{85}, 5639-5642 (2000).

\bibitem{AB-PRA70}A.~Biswas and G.~S.~Agarwal, ``Transfer of an unknown quantum state, quantum networks,
 and memory,'' Phys. Rev. A \textbf{70}, 022323 (2004).

\bibitem{JFZ-PRA73}J.~F.~Zhang, X.~H.~Peng, and D.~Suter, ``Speedup of quantum-state transfer by three-qubit interactions:
Implementation by nuclear magnetic resonance,'' Phys. Rev. A
\textbf{73}, 062325 (2006).

\bibitem{HW-PRA76}H.~Wei, Z.~J.~Deng, X.~L.~Zhang, and M.~Feng, ``Transfer and teleportation of quantum
states encoded in decoherence-free subspace,'' Phys. Rev. A
\textbf{76}, 054304 (2007).

\bibitem{AB-PRA75}A.~Bayat and V.~Karimipour, ``Transfer of d-level quantum states through
spin chains by random swapping,'' Phys. Rev. A \textbf{75}, 022321
(2007).

\bibitem{CDF-PRA81}C.~D.~Franco, M.~Paternostro, and M.~S.~Kim, ``Quantum state transfer via temporal
 kicking of information,'' Phys. Rev. A \textbf{81}, 022319 (2010).

\bibitem{BC-PRA83}B.~Chen, W.~Fan, and Y.~Xu, ``Adiabatic quantum state transfer in a
nonuniform triple-quantum-dot system,'' Phys. Rev. A \textbf{83}, 014301 (2011).

\bibitem{PBL-PRA79}P.~B.~Li, Y.~Gu, Q.~H.~Gong, and G.~C.~Guo, ``Quantum-information transfer
 in a coupled resonator waveguide,'' Phys. Rev. A \textbf{79}, 042339 (2009).

\bibitem{JIC-PRL78}J.~I.~Cirac, P.~Zoller, H.~J.~Kimble, and H.~Mabuchi, ``Quantum state transfer
 and entanglement distribution among distant nodes in a quantum network,'' Phys. Rev. Lett. \textbf{78}, 3221-3224 (1997).

\bibitem{ADB-PRL98}A.~D.~Boozer, A.~Boca, R.~Miller, T.~E.~Northup, and H.~J.~Kimble, ``Reversible
state transfer between light and a single trapped atom,'' Phys. Rev. Lett. \textbf{98}, 193601 (2007).





\bibitem{SB-PRL91}S.~Bose, ``Quantum communication through an unmodulated spin chain,'' Phys. Rev. Lett. \textbf{91}, 207901 (2003).






\bibitem{CHB-PRL70}C.~H.~Bennett, G.~Brassard, C.~Cr$\acute{\texttt{e}}$peau, R.~Jozsa, A.~Peres,
and W.~K.~Wootters, ``Teleporting an unknown quantum state via dual
classical and Einstein-Podolsky-Rosen channels,'' Phys. Rev. Lett.
\textbf{70}, 1895-1899 (1993).

\bibitem{DB-NAT390}D.~Bouwmeester, J.~W.~Pan, K.~Mattle, M.~Eibl, H.~Weinfurter,
and A.~Zeilinger, ``Experimental quantum teleportation,'' Nature
(London) \textbf{390}, 575-579 (1997).

\bibitem{DB-PRL80}D.~Boschi, S.~Branca1, F.~D.~Martini, L.~Hardy, and S.~Popescu, ``Experimental realization of teleporting
 an unknown pure quantum state via dual classical and einstein-podolsky-rosen channels,'' Phys. Rev. Lett. \textbf{80}, 1121-1125 (1998).

\bibitem{MAN-NAT396}M.~A.~Nielsen, E.~Knill, and R.~Laflamme, ``Complete quantum teleportation using
nuclear magnetic resonance,'' Nature (London) \textbf{396}, 52-55 (1998).

\bibitem{yang}C.~P.~Yang, ``Quantum information transfer with superconducting flux qubits coupled to a resonator,'' Phys. Rev. A \textbf{82}, 054303 (2010).






\bibitem{11PLA275}P.~Facchi, V.~Gorini, G.~Marmo, S.~Pascazio, and E.~C.~G.~Sudarshan, ``Quantum
 Zeno dynamics,'' Phys. Lett. A \textbf{275}, 12-19 (2000).

\bibitem{12PRA65}P.~Facchi, S.~Pascazio, A.~Scardicchio, and L.~S.~Schulman, ``Zeno dynamics yields ordinary
constraints,'' Phys. Rev. A \textbf{65}, 012108 (2001).

\bibitem{13PIO42}P.~Facchi and S.~Pascazio, ``Quantum Zeno and inverse quantum Zeno effects,'' Progress in Optics, \textbf{42}, 147-217 (2001).

\bibitem{09JMP18}B.~Misra and E.~C.~G.~Sudarshan, ``The Zeno's paradox in quantum theory,'' J. Math. Phys. \textbf{18}, 756-763 (1977).






\bibitem{PJ-PRL89}P.~Jiannis and W.~Herbert, ``Quantum computation with trapped ions in an optical
 cavity,'' Phys. Rev. Lett. \textbf{89}, 187903 (2002).

\bibitem{KP-PRA69}K.~P.~Jiannis and B.~Almut, ``Decoherence-free dynamical and geometrical entangling
 phase gates,'' Phys. Rev. A \textbf{69}, 033817 (2004).

\bibitem{XQ-PLA374}X.~Q.~Shao, H.~F.~Wang, L.~Chen, S.~Zhang, and K.~H.~Yeon, ``One-step implementation of
 the Toffoli gate via quantum Zeno dynamics,'' Phys. Lett. A \textbf{374}, 28-33 (2009).

\bibitem{SZ-PBA44}S.~Zhang, X.~Q.~Shao, L.~Chen, Y.~F.~Zhao, and K.~H.~Yeon, ``Robust $\sqrt{\texttt{swap}}$ gate on nitrogen-vacancy centres via quantum Zeno
dynamics,'' J. Phys. B: At. Mol. Opt. Phys. \textbf{44}, 075505 (2011).

\bibitem{14PRA77}X.~B.~Wang, J.~Q.~You, and F.~Nori, ``Quantum entanglement via two-qubit quantum
Zeno dynamics,'' Phys. Rev. A \textbf{77}, 062339 (2008).

\bibitem{XQS-EPL90}X.~Q.~Shao, L.~Chen, S.~Zhang, Y.~F.~Zhao, and
K.~H.~Yeon, ``Deterministic generation of arbitrary multi-atom
symmetric Dicke states by a combination of quantum Zeno dynamics and
adiabatic passage,'' Europhys. Lett. \textbf{90}, 50003 (2010).

\bibitem{XQS-NJP12}X.~Q.~Shao, H.~F.~Wang, L.~Chen, S.~Zhang,
Y.~F.~Zhao, and K.~H.~Yeon, ``Converting two-atom singlet state into
three-atom singlet state via quantum Zeno dynamics,'' New J. Phys.
\textbf{12}, 023040 (2010).

\bibitem{ALW-OC284}A.~L.~Wen, ``Distributed qutrit-qutrit entanglement via quantum Zeno
dynamics,'' Opt. Commun. \textbf{284}, 2245-2249 (2011).


\bibitem{15PRA83}A.~L.~Wen and Y.~H.~Guang, ``Deterministic generation of a three-dimensional
 entangled state via quantum Zeno dynamics,'' Phys. Rev. A \textbf{83}, 022322 (2011).





\bibitem{PF-JPC196}P.~Facchi, G.~Marmo, and S.~Pascazio, ``Quantum Zeno dynamics and quantum
Zeno subspaces,'' J. Phys: Conf. Ser. \textbf{196}, 012017 (2009).

\bibitem{PF-PRL89}P.~Facchi, S.~Pascazio, ``Quantum Zeno subspaces,'' Phys. Rev. Lett. \textbf{89}, 080401 (2002).





\bibitem{TP-PRL79}T.~Pellizzari, ``Quantum networking with optical fibres,'' Phys. Rev. Lett. \textbf{79}, 5242 (1997).

\bibitem{AS-PRL96}A.~Serafini, S.~Mancini, ans S.~Bose, ``Distributed quantum computation via optical fibers,'' Phys. Rev. Lett. \textbf{96}, 010503 (2006).




\bibitem{SM-PRA71}S.~M.~Spillane, T.~J.~Kippenberg, K.~J.~Vahala, K.~W.~Goh, E.~Wilcut, and H.~J.~Kimble, ``Ultrahigh-Q toroidal
microresonators for cavity quantum electrodynamics,'' Phys. Rev. A \textbf{71}, 013817 (2005).

\bibitem{JR-PRA67}J.~R.~Buck and H.~J.~Kimble, ``Optimal sizes of dielectric microspheres for cavity QED
with strong coupling,'' Phys. Rev. A \textbf{67}, 033806 (2003).


\bibitem{KJG-IEEE40}K.~J.~Gordon, V.~Fernandez, P.~D.~Townsend, and G.~S.~Buller, ``A short wavelength gigahertz
clocked fiber optic quantum key distribution system,'' IEEE J.
Quantum Electron. \textbf{40}, 900-908 (2004).

\bibitem{FD-PRA75}F.~Dimer, B.~Estienne, A.~S.~Parkins, and H.~J.~Carmichael, ``Proposed realization of the Dicke-model
quantum phase transition in an optical cavity QED system,'' Phys. Rev. A \textbf{75}, 013804 (2007).



\end{thebibliography}
 \end{document}